**Intrinsically Honeycomb-patterned Hydrogenated Graphene with Spin Polarized Edge-states**

*Yang Song, Kai Qian, Lei Tao, Zhenyu Wang, Hui Guo, Hui Chen, Shuai Zhang, Yu-Yang Zhang, Xiao Lin\*, Sokrates T. Pantelides, Shixuan Du\*, Hong-Jun Gao\**

Y. Song, Dr. K. Qian, Dr. L. Tao, Z. Y. Wang, Dr. H. Guo, Prof. H. Chen, Dr. S. Zhang, Prof. Y.-Y. Zhang, Prof. X. Lin, Prof. S. T. Pantelides, Prof. S. X. Du, Prof. H.-J. Gao
Institute of Physics and University of Chinese Academy of Sciences, Chinese Academy of Sciences, Beijing 100190, China
E-mail: E-mail: xlin@ucas.ac.cn; sxdu@iphy.ac.cn; hjgao@iphy.ac.cn

Y. Song, Dr. K. Qian, Dr. L. Tao, Z. Y. Wang, Dr. H. Guo, Prof. H. Chen, Dr. S. Zhang, Prof. Y.-Y. Zhang, Prof. X. Lin, Prof. S. X. Du, Prof. H.-J. Gao
Beijing National Laboratory for Condensed Matter Physics, Beijing 100190, China

Prof. Y.-Y. Zhang, Prof. X. Lin, Prof. S. X. Du, Prof. H.-J. Gao
CAS Center for Excellence in Topological Quantum Computation, Beijing 100190, China

Prof. Y.-Y. Zhang, Prof. X. Lin, Prof. S. X. Du, Prof. H.-J. Gao
Songshan Lake Materials Laboratory, Dongguan 523808, China

Prof. S. T. Pantelides
Department of Physics and Astronomy and Department of Electrical Engineering and Computer Science, Vanderbilt University, Nashville, TN 37235, USA

Since the advent of graphene ushered the era of two-dimensional materials, many forms of hydrogenated graphene have been reported, exhibiting diverse properties ranging from a tunable band gap to ferromagnetic ordering. Patterned hydrogenated graphene with micron-scale patterns has been fabricated by lithographic means. Here we report successful millimeter-scale synthesis of an intrinsically honeycomb patterned form of hydrogenated graphene on Ru(0001) by epitaxial growth followed by hydrogenation. Combining scanning tunneling microscopy observations with density-functional-theory (DFT) calculations, we reveal that an atomic-hydrogen layer intercalates between graphene and Ru(0001). The result is a hydrogen honeycomb structure that serves as a template for the final hydrogenation, which converts the graphene into graphane only over the template, yielding honeycomb-patterned hydrogenated graphene (HPHG). In effect, HPHG is a form of patterned graphane. DFT calculations find that the unhydrogenated graphene regions embedded in the patterned graphane exhibit spin-polarized edge states. This type of growth mechanism provides new pathways for the fabrication of intrinsically patterned graphene-based materials.

## 1. Introduction

Hydrogenation is an effective method to functionalize graphene, as it brings about changes in the electronic structure through the transformation of C-C bonds from $sp^2$ to $sp^3$ hybridization and leads to novel properties. In 2009, a crystalline, fully hydrogenated graphene, named graphane, was successfully fabricated on $SiO_2$.[1] It was later found by theoretical investigations that graphane is an insulator with a large band gap of ~5.4 eV,[2-4] On the other hand, a form of semihydrogenated graphene, named graphone, was theoretically predicted to be a ferromagnetic semiconductor with a small indirect gap[5] and was later fabricated.[6] One-third-hydrogenated graphene (OTHG) has also been fabricated and found to exhibit anisotropic electronic properties.[7]

By now, a large assortment of hydrogenated forms of graphene (HGr) with different carbon-to-hydrogen (C/H) ratios have been predicted[2, 5, 8] or fabricated[7, 9-20]. It has been found that the C/H ratio can be used to tune the band gap.[9, 17, 21] Giant local enhancement of spin-orbit coupling has been predicted [22] and demonstrated when introducing small amounts, ~0.01-0.05%, of hydrogen atoms.[12] Despite the many promising results, however, fabrication of large-area, high-quality HGr with a single C/H ratio remains difficult as growth methods typically lead to a mix of C/H ratios.[13-14, 23] Only micron-scale graphane and millimeter-scale OTHG have been reported so far beyond the nanoscale.[1, 7]

Lithography has been used to fabricate large-area "patterned hydrogenated graphene" (PHG), comprising alternating domains or stripes of pristine graphene and a form of HGr,[24-25] but the pattern dimensions are 100 nm to microns. Nonlithographic nanoscale patterning has been achieved by "templated adsorption" of H on the moiré superlattice of graphene on an Ir substrate.[26] Theoretical studies of this and other nanoscale H patterns on graphene have revealed interesting properties, such as energy-gap scaling[27] and ferromagnetic ordering[28]. On the other hand, in the last few years, a form of "intrinsic patterning" by a dosing-and-annealing process has been demonstrated experimentally in the case of 2D transition-metal chalcogenides, featuring nanoscale 2D triangular patterns and 1D patterns.[29-31] Such a process has not been investigated so far to achieve intrinsically patterned HGr.

In the present work, we successfully synthesize millimeter-scale, intrinsically patterned hydrogenated graphene featuring a honeycomb graphane/graphene pattern on Ru(0001) by directly dosing hydrogen gas on the highly-ordered moiré pattern of graphene on Ru(0001) (Gr/Ru), followed by annealing. Low-energy-electron diffraction (LEED) combined with scanning tunneling microscopy (STM) images show that the honeycomb-patterned hydrogenated graphene (HPHG) contains domains of pristine graphene in the form of hexagonal "holes" surrounded by hydrogenated graphene. The honeycomb structure is quasi-periodic as the diameters of the holes are 2.0 ± 0.2 nm. Combining STM observations with density-functional-theory (DFT)-based calculations, we find that the hydrogenated region is graphane, i.e., HPHG is a form of intrinsically patterned graphane. Moreover, the formation of HPHG comprises two steps: intercalation and hydrogenation. In the intercalation step, hydrogen atoms intercalate at the interface between graphene and Ru substrate and bond to the substrate preferentially under the atop and fcc regions of the moiré pattern of Gr/Ru. As the hydrogen coverage increases, the intercalated hydrogen atoms fill the regions under the atop and fcc regions, producing a hydrogen buffer layer in the form of a nanoscale honeycomb network. In the hydrogenation step, DFT calculations demonstrate that the hydrogen buffer layer serves as a template: hydrogen atoms adsorb preferentially on both sides of the graphene monolayer, but only over the hydrogen honeycomb network, transforming the *sp²* C-C bonds above the honeycomb network to *sp³* C-C bonds, i.e., converting a corresponding honeycomb network of the graphene into graphane. The net result is *intrinsically patterned graphane*: graphane in the form of a honeycomb pattern featuring bare hexagonal graphene regions (the original hcp regions of the moiré pattern). It is noteworthy that the bare hexagonal graphene regions with zigzag edges exhibit spin-polarized edge states similar to the zigzag-edged graphene nanoribbons. This growth mechanism, in which the hydrogen-intercalated layer serves as a template for the growth of HPHG, provides a new pathway for the fabrication of intrinsically patterned graphene-based materials.

## 2. Results and discussion

The first step in the process is the epitaxial growth of monolayer graphene (MLG) on a clean Ru(0001) substrate.[32-33] Gr/Ru exhibits a moiré pattern with a periodicity of 2.91 nm due to the lattice mismatch between graphene and the Ru(0001) surface (**Figure 1**a). Four different regions, namely, atop, bridge, fcc, and hcp regions, can be distinguished in each unit cell of this moiré pattern as shown in the zoom-in view in **Figure 1**d.[7, 33-36] In the next step, HPHG is successfully fabricated through several cycles of exposure of the Gr/Ru system to atomic hydrogen generated by a radio-frequency (RF) atom source at ~200°C, followed by annealing at ~850°C. Here, we define 50 minutes exposure to hydrogen and three hours annealing as one cycle of sample treatment. This choice of cycles is distinctly different from the choice that leads to the fabrication of OTHG.[7]

**Figures 1**b and **1**c show STM images obtained after two and five cycles of sample treatment. Zoom-in images of the black squares in **Figures 1**b and **1**c are shown in **Figures 1**e and **1**f, respectively. At the end of the second cycle, all atop regions appear brighter than in **Figures 1**a and **1**d, indicating the presence of adsorbed hydrogen. In addition, bridge and fcc regions start to be hydrogenated, forming the bright Y-shape patterns seen in **Figures 1**b and **1**e, while the hcp regions are avoided, which gradually becomes the key feature for the formation of HPHG. Each bright Y-shape pattern comprises one hydrogenated fcc region, three neighboring hydrogenated bridge regions, and three hydrogenated atop regions.

After five cycles of sample treatment, all bridge and fcc regions are hydrogenated and a honeycomb network appears, marking the end-point of the HPHG fabrication process (**Figure 1**c,f). The hexagonal "holes" in the pattern are the hcp regions of the original moiré pattern, which continue to be avoided by hydrogen. LEED experiments on different positions of the whole sample show similar diffraction patterns, indicating that HPHG is millimeter-scale and high quality (**Figure S**1).

At this point, all we know about HPHG is that the top side of graphene is hydrogenated in all but the hcp regions of the moiré pattern, giving rise to the honeycomb pattern seen in **Figures 1**c and **1**f. In order to elucidate the formation mechanism and the complete structure of HPHG, we exposed a Gr/Ru sample to atomic hydrogen at a low coverage and annealed at a relatively low temperature (720 °C). A large-area STM image (**Figure 2**a) exhibits bright, imperfect, and discontinuous honeycomb networks, which are reminiscent of the HPHG shown in **Figures 1**c and **1**f, and dark areas. When seen in the zoom-in STM image of **Figure 2**b, the dark areas feature vague patterns that can be seen in the zoom-in STM image shown in **Figure 2**b, which remain to be interpreted.

The line profile along the blue line F in **Figure 2**b, shown in **Figure 2**c, features a small bump in the dark region and a large bump in the bright network. The height of the bright network relative to the small bump is ~1.5 Å (**Figure 2**c). We examined a total of nine line-profiles along the blue lines marked in **Figure 2**b. The heights of the bright networks, which are summarized in **Figure 2**d, exhibit an average value of 1.5±0.1 Å (all the line-profiles are shown in **Figure S2**).

Since the bright features in **Figures 2**a and **2**b correspond to hydrogenated graphene, we built three different models, shown in **Figures 2**e, **2**f, and **2**g to explore the origin of the vague patterns in the dark regions of **Figures 2**b. In each case, the structure is fully relaxed by a DFT calculation. In the model of **Figure 2**e, the height between the hydrogenated and bare graphene areas is ~2.8±0.2 Å, which is much larger than the height measured in the STM images. In view of the fact that intercalation of heteroatoms at the Gr-Ru interface has been reported,[37-38] in the model of **Figure 2**f we assume the presence of an intercalated atomic-hydrogen layer, adsorbed on the Ru surface. Now the height between the hydrogenated and bare graphene areas is ~1.8±0.2 Å (**Figure 2**f), which is in good agreement with the heights measured in the STM image shown in **Figure 2**d. Since the latter are measured from the top of the small bump, we are led to the model shown in **Figure 2**g. The line profile along the red line in **Figure 2**g is presented in **Figure 2**h. The height difference $h'$ indicated in **Figure 2**h is ~1.8±0.2 Å, which is the same as that in **Figure 2**f, but now we reproduce the small bump that is present in the experimental line profiles. We are, therefore, able to interpret the vague patterns seen in the dark areas of **Figure 2**b as a network of intercalated hydrogen atoms adsorbed on the Ru surface, which suggests that the network serves as a template for the formation of the final HPHG structure. The small discrepancy between the experimental value extracted from Figure 2b and the **Figure-2**g model value is likely to be due to the fact that the intercalated-hydrogen networks of **Figure 2**b form at a relatively low annealing temperature and therefore do not correspond to the fully formed and equilibrated networks that are likely to undergird the HPHG. These experimental results and model analysis will now serve as a cornerstone on which we construct a theory for the detailed structure of the HPHG and the atomic-scale mechanism that leads to its formation.

As a first step, we examine the ability of hydrogen atoms to penetrate graphene and intercalate at the Gr/Ru interface. The calculated energy barriers for a H atom to penetrate graphene through hexagons in the atop, bridge, fcc, and hcp regions of the moiré pattern and adsorb on the Ru surface are 1.69 eV, 2.59 eV, 2.82 eV and 2.98 eV, respectively. We infer that, at the annealing temperature of ~850°C, H atoms can most easily penetrate graphene in the atop regions. We then calculated the lowest-energy states of different numbers of H atoms intercalated in the Gr/Ru interface under different regions of the moiré pattern (atop, fcc, bridge, and hcp). We first considered only one intercalated H atom under different regions. For a single H atom under the atop region, the energy is lowest when the H atom adsorbs on the Ru surface. For comparison, if we place the H atom on the Ru surface under the fcc, bridge, and hcp regions, the total energy increases by 0.29 eV, 0.36 eV and 1.73 eV, respectively, suggesting that intercalated H atoms are likely to gradually cover all but the hcp regions (**Figure S4**). Indeed, total energy calculations for configurations with 25, 46 and 52 H atoms adsorbed on the Ru substrate under different regions further confirm that the configurations with H atoms at atop regions, atop/bridge regions, and atop/bridge/fcc regions are always the most energy-favorable configurations. **Figures 3**a-d show the most energy favorable configurations considering adsorption of 1, 25, 46 and 52 H atoms, respectively. The adsorption of H atoms at hcp regions always have a much higher total energy no matter how many H atoms we consider (see **Figure S4-S7** for detailed information).

In order to draw final conclusions about the distribution of intercalated H atoms in the various regions of the moiré pattern, we examine the energy barriers for H adsorbed under the atop regions to diffuse to the other regions. From **Figures 1**d and **3**a, we notice that hcp regions are surrounded by atop and bridge regions, while fcc regions are surrounded by bridge regions. We, therefore, calculated the diffusion barriers of adsorbed H atoms crossing the boundaries between different regions. We find that the H diffusion barriers within and crossing atop, bridge and fcc regions are less than 0.3 eV (**Figures 3**g and **S8**), suggesting that, at the annealing temperature, intercalated H atoms would cover all these regions if sufficient H atoms are provided. The hcp region, however, is very different. In **Figure 3**g we show the total energy variation for a diffusing H atom from the center of an atop region to the center of the adjacent hcp region along the path shown in black in **Figure 3**f and compare it with the corresponding results for H diffusion from the center of the atop region to the center of the fcc region along the path shown in blue in **Figure 3**f. It is clear overall that, though H atoms can easily diffuse and fill the atop, fcc, and bridge regions, they face a steep rise in energy, a barrier of 1.45 eV, to enter the hcp region. Though this barrier can be overcome at the annealing temperature, the duration of the annealing is evidently key to keeping the hcp regions free of intercalated H atoms and forming a honeycomb network (**Figure 3**e). We thus conclude that H atoms intercalate through atop regions and stay on the Ru substrate at atop regions or diffuse to bridge and fcc regions, but not hcp regions, resulting in a honeycomb network at the graphene-Ru interface (**Figure 3**e).

We note that the STM image in **Figure 2**a exhibits imperfect honeycomb bright networks and vague patterns that are somewhat different from the honeycomb pattern shown in **Figure 1**c. The reason is that the STM image shown in **Figure 2** is obtained after annealing at a temperature lower than the prescribed temperature for the proper formation of HPHG. By introducing more H atoms and increasing the annealing temperature, this intermediate state transforms into ideal HPHG.

Once H atoms cover all the atop, bridge and fcc regions at the Gr/Ru interface, further calculations show that the total energy of 13 H atoms adsorbing on Ru under hcp regions is 5.61 eV higher than that of the 13 H atoms hydrogenating the graphene layer (**Figure S9**a,e). Therefore, the experimental and theoretical results so far indicate that the growth mechanism of HPHG on Ru(0001) comprises two steps. In the first step, H atoms intercalate between graphene and the Ru substrate and form a honeycomb network at the interface (**Figure 4**a,I). The second step is the hydrogenation of graphene itself in a way that is guided by the intercalated H on the Ru substrate acting as a template. In order to test if graphene is hydrogenated on both sides, we calculated the energy barriers for H penetration of graphene as it sits on top of the intercalated-H honeycomb template. We found that the energy barrier is 1.6 eV through the atop, bridge and fcc regions, but 2.8 eV through the hcp regions. The net result is that, at the high annealing temperature of 850°C, both sides of graphene can be hydrogenated to form a graphane-like structure, except the hcp regions where H penetration is unlikely.

As a further test to confirm that double-sided hydrogenation of the graphene is energetically favored and to determine which regions hydrogenate first, we consider hydrogenation of the graphene layer using 13 and 43 H atoms at different regions and compare their total energies (**Figure S9** and **S10**). Total-energy calculations of four possible configurations show that upper-side-only hydrogenation is more than 10 eV higher in energy than double-sided hydrogenation (7 and 6 H atoms at the top and bottom side, respectively) at either atop or fcc regions and 5.6 eV higher if all 13 H atoms are placed on the Ru substrate at hcp regions (**Figures S9**a-e). The total energy of the hydrogenated configuration at atop regions on both sides is the lowest, while the relative total energy at fcc regions is 0.73 eV higher. Furthermore, as we increase the number of H atoms from 13 to 43 (**Figure S10**), the total energy of the configuration with double-sided hydrogenation at atop region (**Figure S10**a) is still the lowest, followed by the configuration with double-sided hydrogenation at fcc region (**Figure S10**b), which is consistent with the observations in the STM experiments (**Figure 1**b,c) that atop regions are hydrogenated first and fcc regions next. After all the atop and fcc regions are hydrogenated, a honeycomb structure is finally established (**Figure 4**d and **4**a stage III).

The entire process is depicted schematically in **Figure 4**a.II and III), with the final product being a honeycomb network of hydrogenated graphene, HPHG/Ru(0001), as shown in **Figures 4b, 4c** and **4d**. The calculated lattice constant of HPHG is $a_2 = b_2 = 2.95$ nm and the diameter of the graphene area is $d_2 = 2.0$ nm, which agree well with those from the experimental STM image (**Figure 4**b), $a_1 = b_1 = 2.91 \pm 0.02$ nm and $d_1 = 2.0 \pm 0.2$ nm. The simulated STM image (**Figure 4**c), based on the proposed configuration, also agrees quite well with the experimental image (**Figure 4**b).

We performed additional calculations to find out why hydrogenation of the upper side of the pristine-graphene hcp regions in HPHG/Ru does not happen. We placed 13 H atoms on the upper side at hcp regions and found that the total energy is only 0.13 eV (**Figure S9**f) higher than the energy of a graphene layer that is hydrogenated on both sides at atop regions (**Figure S9**a). This result suggests that the upper sides of hcp regions ought to be hydrogenated as well. A closer look, however reveals that something else is going on. **Figure S11** shows the energy barriers for hydrogen desorption from the upper sides of different regions of HPHG and, for comparison, from freestanding graphane. We found that the H desorption energy barriers for both graphane and HPHG/Ru are ~5 eV, while that for H on the upper side of HPHG hcp regions is only 2.7 eV. The different barriers suggest that the hcp regions, the "holes" in HPHG/Ru, are cleaned during the annealing process so that they end up with no H on either side and the HPHG/Ru is as stable as graphane.

Considering that the one-dimensional (1D) edge states of graphene nanoribbons with zig-zag edges have unique and unusual magnetic structure,[39-40] we performed spin polarized calculations on a freestanding HPHG using the supercell shown in **Figure 4**e. We considered four spin states (**Figure S12**). The total energy of the antiferromagnetic state 1 (AFM-1) is the lowest. The total energy difference per supercell ranging from 109.3 meV to 486.0 meV for different spin states. Similar with zig-zag edged nanoribbons,[40] the total density of states of AFM-1 does not show magnetic moments, but exhibits spin-polarized edge states (**Figure S13**). The spin density distributes along two zig-zag edges on opposite sides of the unhydrogenated graphene region with opposite spin directions (**Figure 4**e). Calculations for a graphene nanoflake with the same structure as the unhydrogenated region in HPHG, with the edges passivated by H, does not show spin-polarized edge states (**Figure S14**), implying that the boundary of bare and hydrogenated graphene plays an important role in forming the spin-polarized edge states in HPHG.

## 3. Conclusion

We have demonstrated a growth mechanism for a newly synthesized HPHG combining STM experiments with DFT calculations. The newly synthesized HPHG, which enriches the HGr family, is a millimeter-scale, highly ordered, nanoscale honeycomb superlattice on a Ru(0001) substrate, a form of intrinsically patterned graphane, and therefore is valuable for basic research and potential applications. The growth process of HPHG comprises two steps. First, a honeycomb intercalated H layer is adsorbed at the Ru surface, guided by the moiré superlattice of Gr/Ru. Second, H atoms adsorb on both sides of graphene at atop, bridge and fcc regions, guided by the intercalated H buffer layer as a template. Furthermore, the intercalated H layer and the HPHG have the same honeycomb pattern, as revealed by STM experiments and DFT calculations. The novel growth mechanism, intercalation-layer-guided hydrogenation, provides a new method for the fabrication of intrinsically patterned graphene-based materials.

## 4. Methods

**Experiments:**

*Preparation of honeycomb-patterned hydrogenated graphene (HPHG) on Ru(0001).* The HPHG was fabricated in a commercial UHV system (Omicron) with a plasma chamber. The system operates with base pressure better than $1\times10^{-10}$ mbar. The Ru(0001) (Mateck) surface was cleaned by repeated cycles of Ar+ sputtering and post annealing at 950 °C. Large-area and high-quality monolayer graphene (MLG) was fabricated by pyrolysis of ethylene on Ru(0001)[32-33]. Ultrahigh vacuum helical inductively coupled RF plasma of 13.56 MHz at a power of 120 W and gas mixture of $H_2$ (15%) and Ar (85%) at a pressure of $1 \times 10^{-4}$ mbar were used. The HPHG was prepared by several cycles of exposure of MLG/Ru(0001) to atomic hydrogen generated by a radio-frequency (RF) atom source. An ion deflection voltage of 250 V was applied to get hydrogen atoms instead of H ions. One cycle refers to exposing the MLG to atomic hydrogen for 50 min at around 200 °C and post annealing at ~850 °C for 3 hours. After preparation, the sample was transferred to an STM chamber and scanned at ~78 K. STM images were acquired in constant-current mode, and all given voltages refer to the sample.

**Calculations:**

Density-functional-theory calculations were performed using projector-augmented wave (PAW) [41-42] pseudopotentials in conjunction with the local-density-approximation (LDA) exchange-correlation functional[43] as implemented in the Vienna Ab-initio Simulation Package (VASP).[44-45] The plane-wave basis was set to an energy cutoff of 400 eV. All the hydrogenated graphene on Ru(0001) systems were modelled by (12 × 12) graphene supercells on a two layered (11 × 11) supercell of Ru(0001) slab (with the lower layer fixed and the upper layer relaxed). The thicknesses of vacuum layers are all larger than 15 Å. All the models were relaxed until the force on each of the relaxed atoms was smaller than 0.05 eV Å$^{-1}$ and the break condition for the electronic self-consistent loop was set at $1 \times 10^{-5}$ eV. The Brillouin zone was sampled by a (1 × 1 × 1) Γ-centered **k**-mesh[46].

We calculated three kinds of energy barriers in this work. First, the penetration barrier of a single H atom passing through Gr in Gr/Ru(0001) system and Gr/52H/Ru(0001) system. Second, the diffusion barrier of a single H atom diffusing on Ru substrate. Third, the desorption barrier of a single H atom desorbing from hydrogenated graphene. These penetration, diffusion and desorption processes were simulated using the climb Nudged Elastic Band (cNEB) method[47-48], with linear interpolation between initial and final states. Before performing cNEB calculations, the initial and final structures were relaxed until the residual force on each of the relaxed atoms was smaller than 0.05 eV Å$^{-1}$. Three intermediate states were constructed by using linear interpolation. In cNEB calculations for penetration barriers, the limit of force convergence was set to 0.1 eV Å$^{-1}$. In the cNEB calculations for diffusion barriers and desorption barriers, the limit of force convergence was set to 0.2 eV Å$^{-1}$.

Different parameters were used in the spin-polarized calculations for freestanding HPHG and the graphene nanoflake. The plane-wave basis was set to an energy cutoff of 600 eV. All the models were relaxed until the force on each atom was smaller than 0.05 eV Å$^{-1}$ and the break condition for the electronic self-consistent loop was set at $1 \times 10^{-6}$ eV. The freestanding HPHG was modelled by its primitive cell, a (12 × 12) graphene supercell with patterned H atoms on it. The same model size as HPHG was used in the calculation of a graphene nano flake. The thicknesses of vacuum layers are all larger than 15 Å. The Brillouin zone was also sampled by a (1 × 1 × 1) Γ-centered **k**-mesh.

**Supporting Information**
Supporting Information is available from the Wiley Online Library or from the author.

**Acknowledgements**
This work was supported by grants from Science Foundation of China (Nos. 51922011 and 61888102), National Key Research and Development Projects of China (Nos. 2016YFA0202300, 2018YFA0305800 and 2019YFA0308500), the Strategic Priority Research Program of the Chinese Academy of Sciences (Nos. XDB28000000 and XDB30000000), the International Partnership Program of Chinese Academy of Sciences (No. 112111KYSB20160061), the K. C. Wong Education Foundation and the Fundamental Research Funds for the Central Universities. Computational resources were provided by the National Supercomputing Center in Tianjin. Work at Vanderbilt was funded by the U.S. Department of Energy, Office of Science, Basic Energy Sciences, Materials Science and Engineering Division grant no. DE-FG02-09ER46554 and by the McMinn Endowment. Y.S., K.Q. and L.T. contributed equally.

Received: ((will be filled in by the editorial staff))
Revised: ((will be filled in by the editorial staff))
Published online: ((will be filled in by the editorial staff))

**Conflict of interest**
The authors declare no conflict of interest.

**Keywords**
2D material, hydrogenated graphene, intercalation layer, ferromagnetic semiconductor

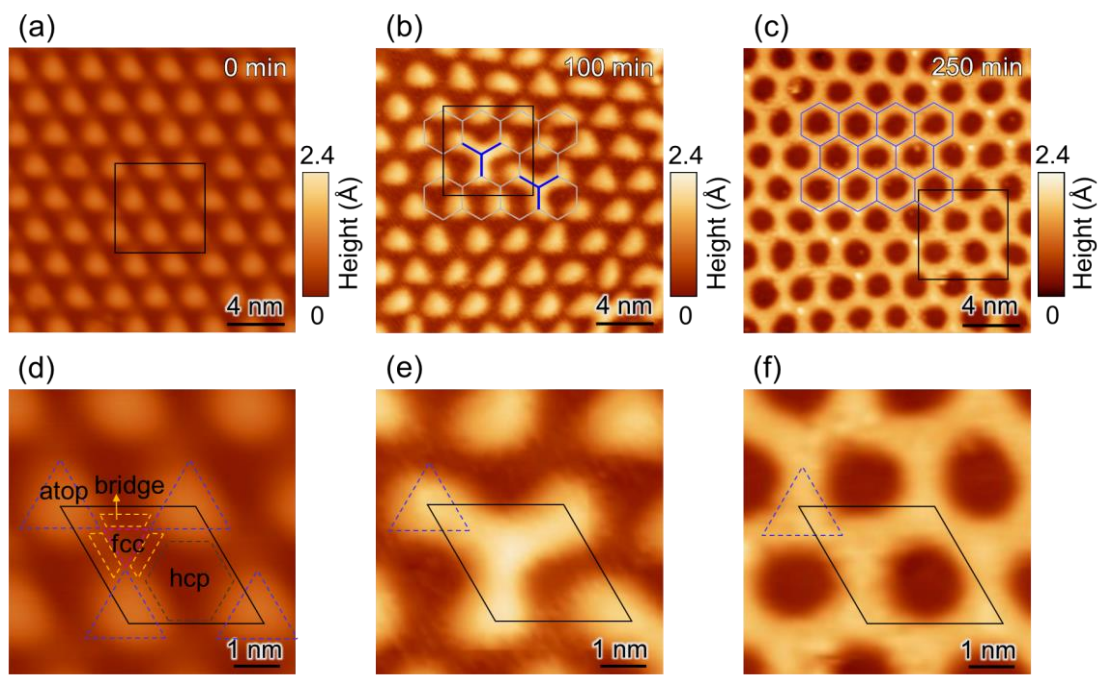

**Figure 1.** STM images of graphene on Ru(0001) with increasing H coverage. (a) STM image of graphene on Ru(0001) (Gr/Ru). The triangular pattern shows the moiré pattern of Gr/Ru with bright atop regions [atop, bridge, fcc, and hcp regions are identified in the zoom-in image of panel (d), corresponding to the area in the black box in panel (a)]. (b,c) STM images of Gr/Ru after exposure to atomic hydrogen for 100 min and 250 min, respectively. In (b) the atop regions are brighter than in (a), indicating hydrogenation. The Y-shaped bright areas, marked by blue "Y", signal the beginning of hydrogenation of fcc and bridge regions [compare panels (d) and (e)], comprising three hydrogenated atop regions and one fcc region surrounded by three bridge regions, while the hcp regions are still bare. (d-f) Zoom-in STM images of the black squares in (a-c), respectively. The black rhombus marks the unit cell of the Gr/Ru moiré pattern. The scanning conditions are: (a) U = − 0.5 V, I = 0.5 nA; (b) U = − 0.5 V, I = 0.6 nA; (c) U = − 1.0 V, I = 0.2 nA.

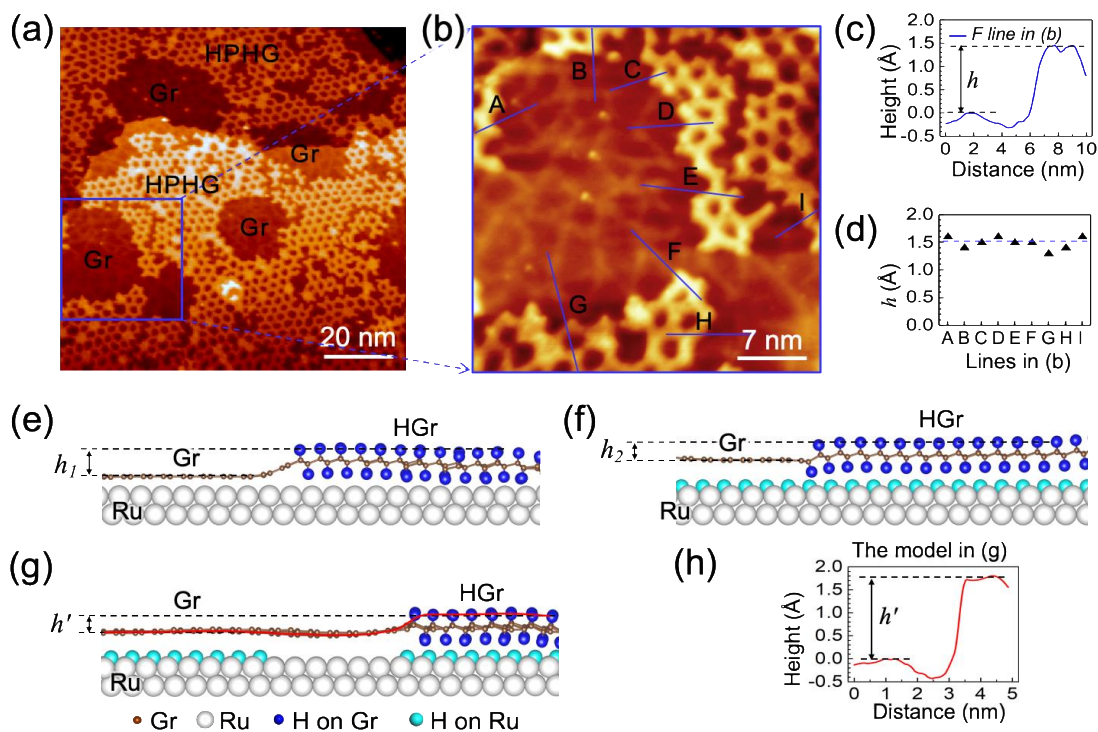

**Figure 2.** STM Images and height profiles of HPHG on Ru(0001) at low H coverage and low annealing temperature. (a) Large scale STM image of mixed structures of graphene and HPHG on Ru(0001). The scanning conditions are U = − 1.5 V, I = 0.05 nA. (b) Zoom-in STM image of the blue square in (a). There are vague patterns in the graphene region, suggesting a hydrogen network is formed under the graphene. (c) Line profile along the blue line F in (b), which crosses Gr/H/Ru and HPHG/H/Ru regions. The height difference of Gr/H/Ru and HPHG/H/Ru, $h$, is 1.5 Å. (d) The statistical distribution of $h$ according to the line profiles along all the blue lines in (b). The mean value of $h$ is 1.5±0.1 Å. (e-g) Side views of three models of the atomic structure of partially hydrogenated graphene on Ru(0001). In (e), the height difference of HGr/Ru and Gr/Ru ($h_1$) is 2.8±0.2 Å. (f) In (f), the height difference of HGr/H/Ru and Gr/H/Ru ($h_2$) is about 1.8±0.2 Å, which agrees well with the experimental observations in (d). In (g), the red profile line, reproduced in (h), reveals a height of 1.8±0.2 Å and also reproduces the small bump seen in the experimental line profiles (Figure 2c and Figure S2). (h) The height profile along the red line in (g). Detailed information is shown in Figure S3. The height difference of Gr/H/Ru and HGr/H/Ru, $h'$, is ~1.8 Å, which is close to the experimental value in (c). The shape of the curve looks similar to that of the experimental line profile in (c).

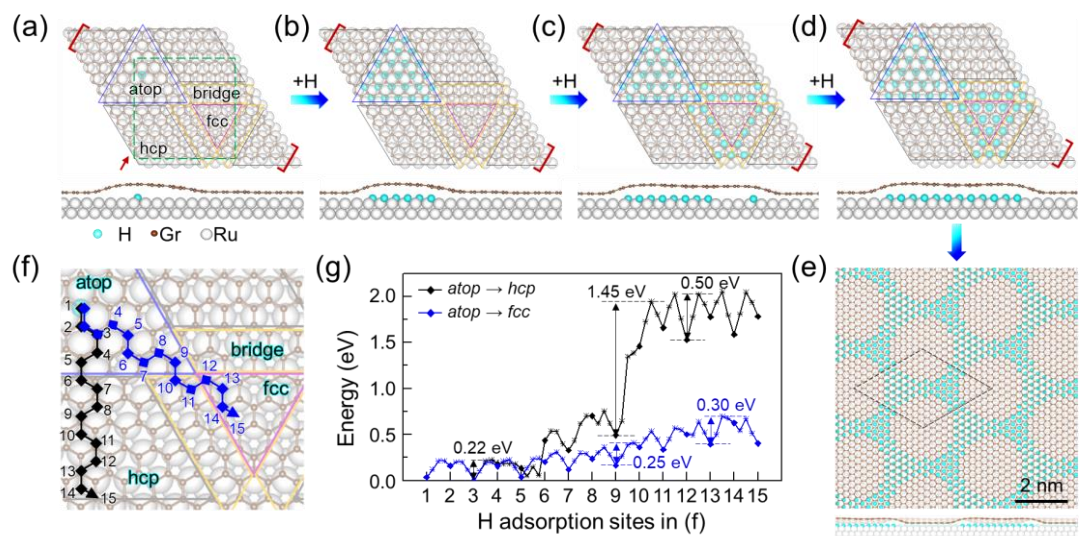

**Figure 3.** DFT calculations on the formation of an intercalated honeycomb H layer at the graphene-Ru(0001) interface. Cyan and blue are used to distinguish H atoms adsorbed on the Ru substrate from H atoms bonded to graphene. (a-d) The growth process of honeycomb intercalated H layer starting with (a) one H atom adsorbing at the Ru surface under an atop region; (b) 25 H atoms adsorbed under an atop region, and (c) 46 H under atop/bridge regions, and (d) 52 H atoms under atop/bridge/fcc regions. The blue triangle, orange trapezoid, deep pink triangle, and gray triangle and parallelogram mark the atop, bridge, fcc and hcp regions, respectively. To make the figures clear, all the side views are looking from the direction marked with red arrow and only contain the atoms in the range marked by the red brackets to show the change of the corrugation in graphene. (e) A zoom-out model of (d) showing a honeycomb intercalated H layer. (f) Zoom-in of the green dashed rectangle in (a). The arrows mark the diffusion paths for a H atom diffusing from the center of the atop region to the center of the hcp region (black arrows) and the center of the fcc region (blue arrows). (g) The black line and the blue line are the diffusion barriers for a H atom diffusing along the routes indicated by black arrows and blue arrows in (f).

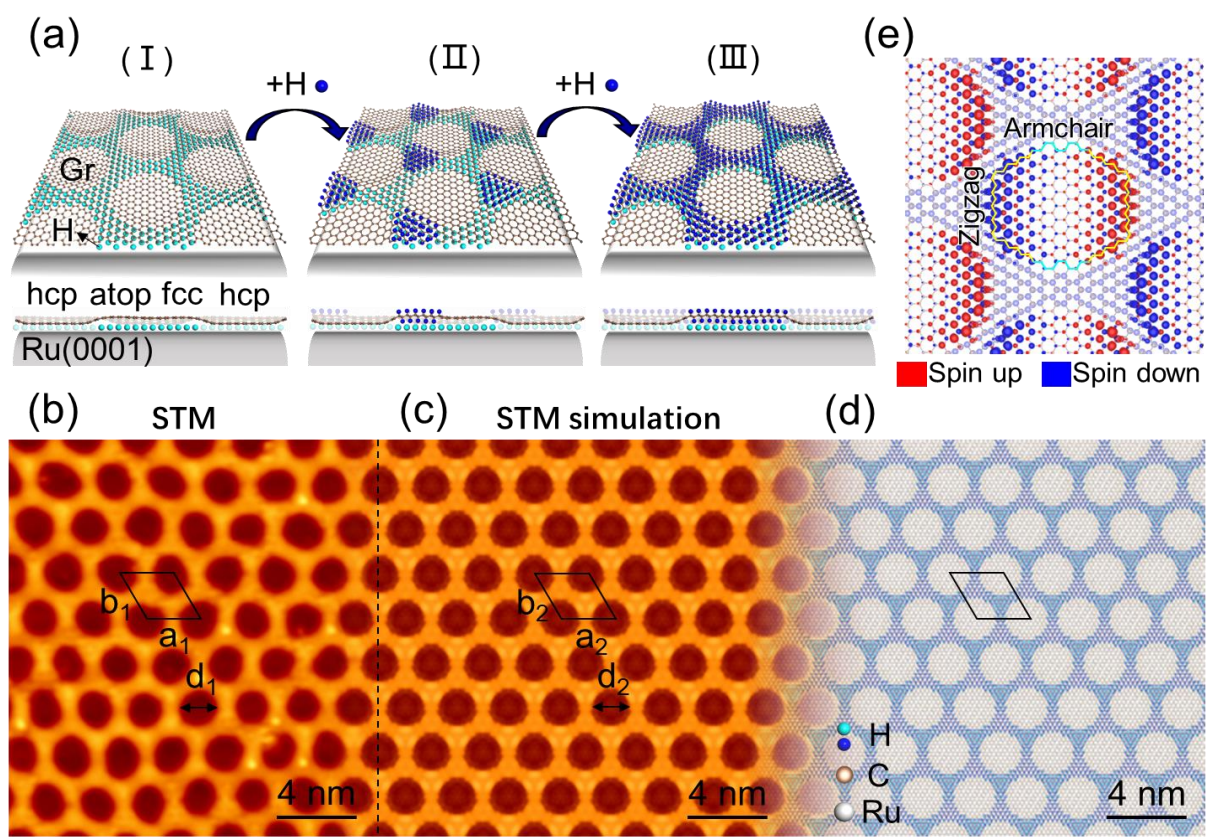

**Figure 4.** Schematic of the growth mechanism of HPHG on Ru(0001). (a) Schematics of the growth mechanism of HPHG/Ru(0001). (I) H atoms form a honeycomb pattern at the Gr/Ru interface by passing through graphene and adsorbing on the Ru(0001) surface to fill up atop, bridge and fcc regions. (II) H atoms adsorb on both sides of graphene at atop regions. (III) H atoms adsorb on both sides of graphene at all atop, bridge and fcc regions. The configurations have been fully relaxed by DFT calculations. (b) Experimental STM image of HPHG/Ru(0001), U = − 0.1 V, I = 0.2 nA. The lattice constant $a_1$ ($b_1$) is 2.91 nm. (c) Simulated STM image of HPHG/Ru(0001) at an energy range from -0.1 eV to Fermi level. The lattice constant is 2.95 nm. (d) Atomic structure of HPHG/Ru(0001) used to do the STM simulation in (c). (e) Top view of a freestanding HPHG shows the spin charge density differences ($\rho\uparrow − \rho\downarrow$) along the edges of the unhygrogenated hcp region. The isosurfaces of $\Delta\rho\uparrow$ and $\Delta\rho\downarrow$ charge density are marked by red and blue, respectively. The spin distribution demonstrates that HPHG is antiferromagnetic. The isosurface is chosen as 0.0016 e bohr$^{−3}$.

TOC:

Intrinsically honeycomb-patterned hydrogenated graphene with millimeter-scale has been successfully synthesized by epitaxial method and the growth mechanism has been revealed by DFT calculations. The growth mechanism is that the intercalated H layer serves as a template for the double-sided hydrogenation of the graphene layer. DFT calculations further reveals that monolayer HPHG is an antiferromagnetic semiconductor.

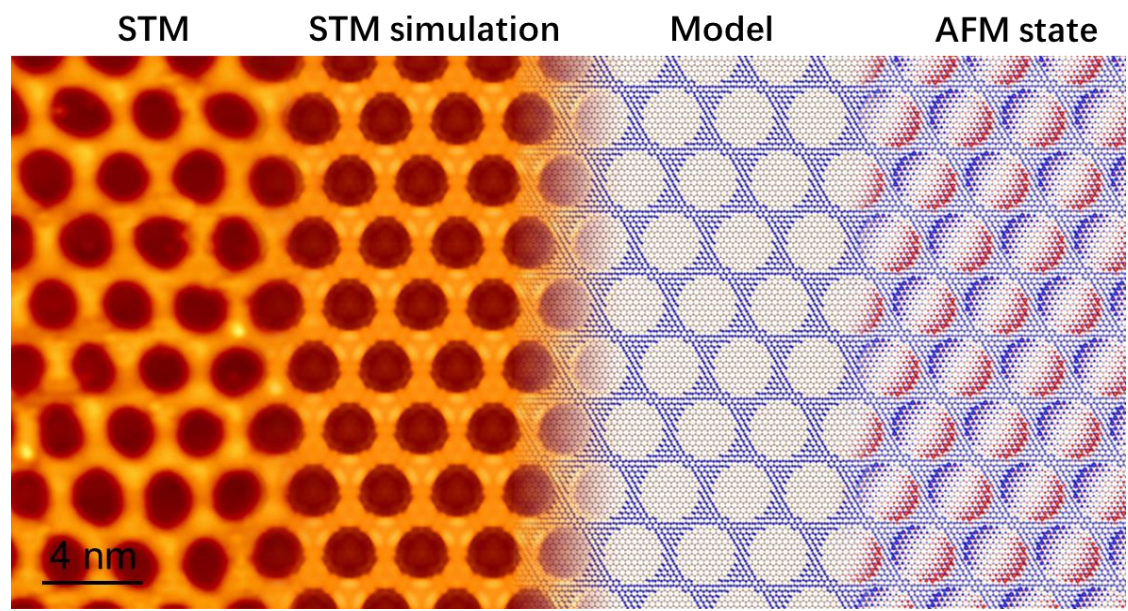